# Generation of non-classical correlated photon pairs via a ladder-type atomic configuration: theory and experiment

Dong-Sheng Ding, Zhi-Yuan Zhou, Bao-Sen Shi,* Xu-Bo Zou,[1] and Guang-Can Guo

[1]*Key Laboratory of Quantum Information, University of Science and Technology of China, Hefei 230026, China*
[2]*xbz@ustc.edu.cn*
**drshi@ustc.edu.cn*

**Abstract:** We experimentally generate a non-classical correlated two-color photon pair at 780 and 1529.4 nm in a ladder-type configuration using a hot $^{85}$Rb atomic vapor with the production rate of ~$10^7$/s. The non-classical correlation between these two photons is demonstrated by strong violation of Cauchy-Schwarz inequality by the factor R = 48 ± 12. Besides, we experimentally investigate the relations between the correlation and some important experimental parameters such as the single-photon detuning, the powers of pumps. We also make a theoretical analysis in detail and the theoretical predictions are in reasonable agreement with our experimental results.

## References and links

## 1. Introduction

Non-classical correlated photon pairs are important photon sources for many applications such as Franson interference [1], quantum ghost imaging [2] and interference [3], quantum communication [4] and quantum teleportation [5], and so on. Usually, entangled photon pairs are generated through spontaneously parametric down-conversion (SPDC) process in a nonlinear crystal [6]. One disadvantage of this way is that the photon generated has so wide bandwidth (~THz) that it can't effectively couple with atom (natural bandwidth: ~MHz). Such coupling is a key for realizing long-distance quantum communication using atoms [4]. So it is very important to generate correlated photons with narrow bandwidth. This problem can be solved by following two ways: the SPDC with the aid of cavity [7–11] and spontaneously Raman scattering (SRS) [12–14] or spontaneously four-wave mixing (SFWM) [15–18] in an atomic system. Although the way based on SRS or SFWM causes high attention, and there are some progresses along this direction [12–18] recently, there is one problem: the wavelengths of the photons generated are not in the transmission window of fiber [12–18], these photons are not suitable for long-distance transmission due to large loss when they transmit in the fiber. Ref. 19 clearly shows that the time needed to establish an entanglement between two legal users could be significantly reduced if a photon in telecom-band is used as an information carrier, compared with the case of a photon in near infrared band used. Therefore it is desirable to prepare a photon whose wavelength is in the transmission window of the fiber. There are some progresses along this direction very recently: Kuzmich et. al. [20] and Willis et. al [21] have prepared non-classical correlated two-color photon pairs in a cold atomic ensemble and a hot atomic vapor respectively. Besides the narrow bandwidth the photon has, the wavelength of one photon of each pair in these two works is in a transmission window of the fiber and the other is at a transition wavelength of the atom. These photons are very useful for the build-up of the long-distance quantum network based on atomic quantum repeater.

In this paper, we consider how to prepare a non-classical correlated two-color photon pair via SFWM through a ladder-type configuration in hot $^{85}$Rb atoms. We make a detail theoretical analysis about this SFWM process by using perturbation theory. After deriving some useful expressions about the relations between the cross-correlation and other interesting parameters, we make some numerical calculations by the parameters used in our experiments. After that we experimentally generate such kind of photon pair, obtain the photon pair production rate of ~$10^7$/s. From experimental results of correlation measurement, we obtain the strong violation of Cauchy-Schwarz inequality by the factor R = 48 ± 12. The frequency of the photon in telecomband corresponds to the $^{85}$Rb atom transition between $5P_{3/2}(F'=3) \rightarrow 4D_{5/2}(F''=4)$, the other photon is at 780 nm, corresponding to the transition $5S_{1/2}(F=3) \rightarrow 5P_{3/2}(F'=3)$. The bandwidth of photon is about 450 MHz. Besides, the relations between the non-classical correlation and some important parameters such as the detuning of the single photon and the powers of pumps are experimentally obtained. The theoretical predictions are in reasonable agreement with the experimental results.

Although our experimental configuration is similar to that in Ref [20], a cold atomic ensemble is used in that work. Therefore our setup is much simpler: there is no need for laser trapping, high vacuum, one more interesting thing is that our system can be miniaturized. Very recently, Willis et. al. [21] obtain correlated photon pairs using a diamond configuration in a hot atomic ensemble. A difference is that a co-propagation configuration is used in their setup, compared with a counter-propagation configuration used instead in our experiment. One more important thing we want to point out is that different configurations used in theirs and our experiments induce very different generation efficiencies: In Ref. 21, the pump power of 795 nm laser used is about 15 mW, is about three orders higher than in our experiment (pump 2: ~10 µW). The photon production rate is about 4500/s in their work [22], much lower compared with ~$10^7$/s production rate in our experiment. Besides, we also investigate the relations between the non-classical correlation and some important experimental parameters, such as the detuning of the single-photon, power of pump laser. Our work is definitely promising to the research field of long-distance quantum communication network based on atomic quantum repeater.

The outlines of this paper are follows: after giving a simple introduction in section 1, we briefly introduce the atomic configuration experimental setup in section 2. In section 3, we give a detail theoretical description. After that we show our experimental results in section 4. Finally, we give a brief conclusion in section 5.

## 2. Experimental setup

The correlated photon pairs are generated using a ladder-type configuration of energy levels consisted of four states |1>, |2>, |3> and |4>, where |2> and |3> are degenerate states. The configuration is shown in Fig. 1. A 5-cm-long cell filled with isotopically pure $Rb^{85}$ is used to produce correlated photon pairs. The four states |1>, |2> (|3>) and |4> correspond to $Rb^{85}$ atomic levels $5S_{1/2}(F=3)$, $5P_{3/2}$ and $4D_{5/2}$ respectively. The states |4> and |2> (|3>) have a

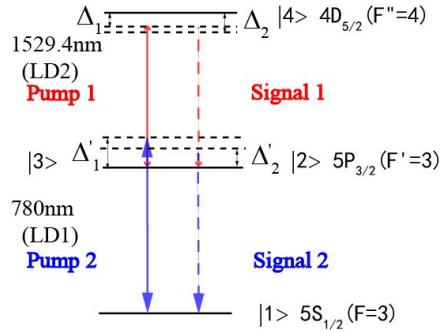

Fig. 1. Energy levels diagram of the ladder-type configuration in our experiment. $\Delta_1$, $\Delta'_1$ is the detuning of the frequency of the Pump 1 and Pump 2 with the definitions $\Delta'_1 = \omega_{p2}-\omega_{31}$, $\Delta_1 = \omega_{43}-\omega_{p1}$. $\Delta_2$, $\Delta'_2$ is the detuning of the frequency of the Signal 1 and Signal 2 with the definitions $\Delta'_2 = \omega_{s2}-\omega_{21}$, $\Delta_2 = \omega_{42}-\omega_{s1}$.

radiative lifetime broadening of 1.7 MHz and 6 MHz respectively. In Fig. 1, the wavelength of Pump 1 is 1529.4 nm, corresponding to the atoms $^{85}$Rb atom transition $5P_{3/2}(F'=3) \rightarrow 4D_{5/2}(F''=4)$, and wavelength of Pump 2 is 780 nm, corresponding to the transition $5S_{1/2}(F=3) \rightarrow 5P_{3/2}(F'=3)$. The wavelengths of the generated Signal 1 photon and Signal 2 photon are 1529.4 nm and 780 nm respectively. $\Delta_1$, $\Delta'_1$ is detuning of the frequency of the Pump 1 and Pump 2 with the definitions of $\Delta'_1 = \omega_{p2}-\omega_{31}$, $\Delta_1 = \omega_{43}-\omega_{p1}$. $\Delta_2$, $\Delta'_2$ is the detuning of the frequency of the Signal 1 and Signal 2 with the definitions of $\Delta'_2 = \omega_{s2}-\omega_{21}$, $\Delta_2 = \omega_{42}-\omega_{s1}$. The red frequency shift of $\Delta$ represents positive in this paper.

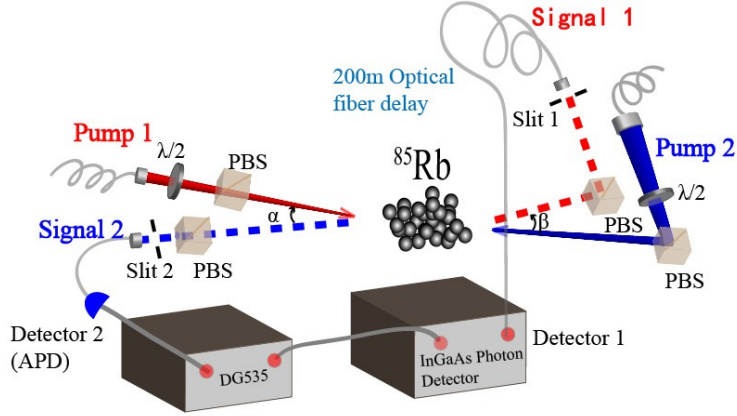

Fig. 2. Experimental setup. Signal 2 photons are collected by a multi-mode fiber, Signal 1 photons are delayed using a 200-m long single-mode fiber firstly, then are collected by a single-mode fiber.λ/2: half-wave plate;PBS: polarization beam splitter; DG535: delay generator; Detector 1: In-GaAs Photon Detector; Detector 2: Avalanche diode; α,β are the angles of the Pump 1 and Signal 2, and the Pump 2 and Signal 1, and α = β = 3.6°; Slit 1 and slit 2 are slits which are used to reduce noise.

Our experimental setup is shown in Fig. 2. A continuous wave (cw) laser at 780 nm from an external-cavity diode laser (DL100, Toptica) is input to the Rb cell as the Pump 2 and a cw laser beam at 1529.4 nm from another external-cavity diode laser (DL Prodesign, Toptica) is the Pump 1. The power of Pump 1 with horizontal polarization is controlled by a half-wave plate and a polarization beam splitter. Similarly, the power of vertically polarized Pump 2 is adjusted through another half-wave plate and another polarization beam splitter. The angle between two pump lights is $0.9°$, and the waists of Pump 1 and Pump 2 at the center of Rb cell are 0.68 mm and 1 mm respectively. The directions of generated correlated photons are defined by the angle $\alpha = \beta = 3.6°$, where α(β) is the angle between Pump 1 and Signal 2 (Pump 2 and Signal 1). We use a single-mode fiber and a multi-mode fiber to collect Signal 1 and Signal 2 photons respectively (The reason why we use a multi-mode fiber to couple Signal 2 is only to improve the trigger rate of the detector 1 by increasing single photon counts of the detector 2. By which the measure time of cross-correlation could be shorten, and the influence from the stability of the experimental system could be reduced.). Signal 2 photon is detected by detector 2 (Avalanche diode, PerkinElmer SPCM-AQR-15-FC with 50% efficiency.). After about 1000 ns time delay caused in a 200-m long fiber, Signal 1 photon is recorded by the detector 1 working at gated mode (ID Quantique, InGaAs Photon Detector with 8% detection efficiency), triggered by a delayed electronic pulse generated by DG535 when the detector 2 fires.

### 3. Theoretical analysis

Before showing our experimental results, we firstly give a detail theoretical description about the SFWM process. We denote the electric field as $E = [\hat{E}^{(+)} + \hat{E}^{(-)}]/2 = [\hat{E}^{(+)} + c.c]/2$. The four fields along z axis shown in Fig. 1 are (The angles between these modes are very small, therefore we regard all the modes near parallel)

$$\hat{E}_{p1}^{(+)} = \varepsilon_{p1} \exp(-i\omega_{p1}t + ik_{p1}z),$$

$$\hat{E}_{p2}^{(+)} = \varepsilon_{p2} \exp(-i\omega_{p2}t - ik_{p2}z),$$

$$\hat{E}_{s1}^{(+)} = \frac{1}{\sqrt{2\pi}} \int d\omega \sqrt{\frac{2\hbar\omega_{s1}}{c\varepsilon_0 A}} \hat{a}_{s1} \exp(-i\omega_{s1}t + ik_{s1}z), \quad (1)$$

$$\hat{E}_{s2}^{(+)} = \frac{1}{\sqrt{2\pi}} \int d\omega \sqrt{\frac{2\hbar\omega_{s2}}{c\varepsilon_0 A}} \hat{a}_{s2} \exp(-i\omega_{s2}t - ik_{s2}z).$$

where, $c$ is velocity of light in vacuum, $\hbar$ is Planck's constant, $\varepsilon_0$ is the permittivity of vacuum, $t$ is the time, $\varepsilon_j (j = p1, p2)$ are the amplitudes of the electromagnetic fields Pump1 and Pump 2. $\omega_i$ and $k_j (j = s1, s2, p1, p2)$ are the frequency and the wave vector (Signal 1, Signal 2, Pump 1 and Pump 2); Pump 1 and Pump 2 can be regarded as the classical fields because they are strong. Signal 1 and Signal 2 are generated fields, and are described by annihilation operators $\hat{a}_{s1}, \hat{a}_{s2}$. The Rabi frequencies of the four fields Pump 1, Pump 2, Signal 1 and Signal 2 are

$$\Omega_{p1} = \frac{u_{34}\hat{E}_{P1}^{(+)}}{\hbar}, \Omega_{p2} = \frac{u_{13}\hat{E}_{P2}^{(+)}}{\hbar}, \Omega_{s1} = \frac{u_{42}\hat{E}_{S1}^{(+)}}{\hbar}, \Omega_{s2} = \frac{u_{21}\hat{E}_{S2}^{(+)}}{\hbar}. \quad (2)$$

where, $\mu_{ij}$ is the electric dipole matrix element related to the transition between levels $i$ and $j$. Assumed the two-photon detuning $\Delta_1'-\Delta_1=\Delta_2'-\Delta_2$, the effective interaction Hamiltonian for this system can be written as

$$\begin{aligned}H_{int} = &-\hbar\Delta_1'|3\rangle\langle 3| - \hbar\Delta_2'|2\rangle\langle 2| + \hbar(\Delta_2 - \Delta_2')|4\rangle\langle 4| \\ &-\hbar(\Omega_{s2}|2\rangle\langle 1| + \Omega_{p2}|3\rangle\langle 1| + \Omega_{p1}|4\rangle\langle 3| + \Omega_{s1}|4\rangle\langle 2| + H.c).\end{aligned} \quad (3)$$

The density matrix equation of motion is

$$\dot{\rho}_{ij} = -\frac{i}{\hbar}\sum_k (H_{ik}\rho_{kj} - \rho_{ik}H_{kj}) - \Gamma_{ij}\rho_{ij}. \quad (4)$$

where, $\Gamma_{ij}$ ($i \neq j$) describes the complex decay rate from $|i\rangle$ to $|j\rangle$, $\Gamma_{ij}$ ($i = j$) describes the decay rate of $\rho_{ii}$. $\Omega_{p1}, \Omega_{p2} \gg \Omega_{s1}, \Omega_{s2}$. In our system, the states $|2\rangle, |3\rangle$ are degenerate, their decay rates are $2\gamma_1$. The decay rate of state $|4\rangle$ is $2\Gamma_1$. The equations of motion describing the evolution of this system can be written as:

$$\dot{\rho}_{21} = i\Omega_{s2}(\rho_{11} - \rho_{22}) + i\Omega_{s1}^*\rho_{41} - i\Omega_{p2}\rho_{23} - \rho_{21}\Gamma_{21} \quad (5)$$

$$\dot{\rho}_{31} = i\Omega_{p2}(\rho_{11} - \rho_{33}) + i\Omega_{p1}^*\rho_{41} - i\Omega_{s2}\rho_{32} - \rho_{31}\Gamma_{31} \quad (6)$$

$$\dot{\rho}_{42} = i\Omega_{s1}(\rho_{22} - \rho_{44}) + i\Omega_{p1}\rho_{32} - i\Omega_{s2}^*\rho_{41} - \rho_{42}\Gamma_{42} \quad (7)$$

$$\dot{\rho}_{43} = i\Omega_{p1}(\rho_{33} - \rho_{44}) + i\Omega_{s1}\rho_{23} - i\Omega_{p2}^*\rho_{41} - \rho_{43}\Gamma_{43} \quad (8)$$

$$\dot{\rho}_{41} = i\Omega_{s1}\rho_{21} + i\Omega_{p1}\rho_{31} - i\Omega_{s2}\rho_{42} - i\Omega_{p2}\rho_{43} - \rho_{41}\Gamma_{41} \quad (9)$$

$$\dot{\rho}_{32} = -i\Omega_{s1}\rho_{34} - i\Omega_{s2}^*\rho_{31} + i\Omega_{p1}^*\rho_{42} + i\Omega_{p2}\rho_{12} - \rho_{32}\Gamma_{32} \quad (10)$$

where, $\Gamma_{21}, \Gamma_{31}, \Gamma_{42}, \Gamma_{43}, \Gamma_{41}$ and $\Gamma_{32}$ are complex relaxation rates defined as: $\Gamma_{21}=\gamma_1-i\Delta_2'$, $\Gamma_{31}=\gamma_1-i\Delta_1'$, $\Gamma_{42}=2\Gamma_1+\gamma_1-i\Delta_2$, $\Gamma_{43}=2\Gamma_1+\gamma_1-i\Delta_1$, and $\Gamma_{41}=2\Gamma_1+i(\Delta_2-\Delta_2')$. We consider the zero-order perturbation expansion with the assumption of $\Omega_{p1} \gg \Omega_{s1}, \Omega_{s2}, \Omega_{p2}$ (In our experiment, $\Omega_{p1} \gg \Omega_{p2}$), and derive the steady-state solutions of the above equations. After some calculations, we obtain the first-order induced atomic susceptibility $\hat{P}_{s2}^{(1)}$, describing the Raman gain of Signal 2 field, and the third-order nonlinear induced atomic susceptibility $\hat{P}_{s2}^{(3)}$, describing the generation of Signal 2 field, respectively

$$\hat{P}_{s2}^{(1)} = \varepsilon_0 \frac{-iNu_{21}^2(\Gamma_{14} + 2\gamma_1)}{\hbar\varepsilon_0(\Gamma_{13}\Gamma_{14} + \Omega_{p1}^2)} \hat{E}_{S2}^{(+)}. \tag{11}$$

$$\hat{P}_{s2}^{(3)} = \varepsilon_0 \frac{-iNu_{21}u_{13}u_{34}u_{42}}{\hbar^3\varepsilon_0\Gamma_{21}(\Gamma_{31}\Gamma_{41} + \Omega_{p1}^2)} \hat{E}_{P1}^{(+)} \hat{E}_{P2}^{(+)} \hat{E}_{S1}^{(-)}. \tag{12}$$

where, $N$ is the atomic active density. From Eq. (11) and (12), we can easily calculate the first-order and third-order nonlinear induced atomic susceptibilities related to the generated Signal 2 field

$$\chi_{s2}^{(1)} = \frac{-iNu_{21}^2(\Gamma_{14} + 2\gamma_1)}{\hbar\varepsilon_0(\Gamma_{13}\Gamma_{14} + \Omega_{p1}^2)}. \tag{13}$$

$$\chi_{s2}^{(3)} = \frac{-iNu_{21}u_{13}u_{34}u_{42}}{\hbar^3\varepsilon_0\Gamma_{21}(\Gamma_{31}\Gamma_{41} + \Omega_{p1}^2)}. \tag{14}$$

The state of two-color correlation photons pairs generated through SFWM can be described [23–25] as

$$|\psi\rangle = L\int d\omega_{s1}d\omega_{s2}\kappa(\omega_{s1},\omega_{s2})\delta(\omega_{p1} + \omega_{p2} - \omega_{s2} - \omega_{s1})\sinc(\frac{\delta kL}{2})a_{s1}^\dagger(\omega_{s1})a_{s2}^\dagger(\omega_{s2})|0\rangle. \tag{15}$$

where, L is the length of cylindrical volume, |0> is vacuum state. $\delta k = k_{p1} - k_{p2} + k_{s1} - k_{s2}$ is the mismatch of these vectors. $a_{s1}^\dagger, a_{s2}^\dagger$ are the creation operators of Signal 1 and Signal 2. $\kappa(\omega_1,\omega_2) = -i(\sqrt{\bar{\omega}_1\bar{\omega}_2}/2c)\chi^{(3)}(\omega_1,\omega_2)\varepsilon_{p1}\varepsilon_{p2}$, here, $\bar{\omega}_1, \bar{\omega}_2$ are the central frequencies of Signal 1 and Signal 2. $\chi^{(3)}(\omega_1,\omega_2)$ is the third-order nonlinear susceptibility related to Signal 1(or Signal 2) field. In our system, the conditions

$$\begin{aligned} k_{p1} - k_{p2} + k_{s1} - k_{s2} &= 0, \\ \omega_{p1} + \omega_{p2} - \omega_{s2} - \omega_{s1} &= 0, \end{aligned} \tag{16}$$

should be satisfied due to the phase matching and energy conservation requirements in SFWM. Then, Eq. (15) becomes

$$|\psi\rangle = L\int d\omega_{s2}\kappa(\omega_{p1} + \omega_{p2} - \omega_{s2},\omega_{s2})\sinc(\frac{\delta kL}{2})a_{s1}^+(\omega_{p1} + \omega_{p2} - \omega_{s2})a_{s2}^+(\omega_{s2})|0\rangle. \tag{17}$$

With similar treatment shown in Ref [23–25], we obtain the function of the two-photon as following:

$$G^{(2)}(\tau) = \left|L\int d\tau' \tilde{\kappa}(\tau')\tilde{\Phi}(\tau - \tau')e^{-i(\omega_c + \omega_p)t_{s1}}\right|^2. \tag{18}$$

where, $\tau = t_{s2} - t_{s1}$ is relative time, $\tilde{\kappa}(\tau')$, $\tilde{\Phi}(\tau - \tau')$ are the Fourier transforms of $\kappa(\omega_{s2})$, $\Phi(\omega_{s2})$ respectively, $\tilde{\kappa}(\tau') = \frac{1}{2\pi}\int d\omega_{s2} \kappa(\omega_{s2}) e^{-i\omega_{s2}\tau'}$, $\tilde{\Phi}(\tau - \tau') = \frac{1}{2\pi}\int d\omega_{s2} \Phi(\omega_{s2}) e^{-i\omega_{s2}(\tau-\tau')}$.

Here, $\kappa(\omega_{s2}) = -i(\sqrt{\overline{\omega}_{s1}\overline{\omega}_{s2}}/2c)\chi^{(3)}(\omega_{s2})\varepsilon_{p1}\varepsilon_{p2}$, $\Phi(\omega_{s2}) = \mathrm{sinc}(\frac{\delta k L}{2})e^{i(k_{s2}+k_{s1})L/2}$.

The relations between the wavenumbers of signal fields and the first-order susceptibility are:

$$k_{s1} = \frac{\omega_1}{c}\sqrt{1+\chi_{s1}^{(1)}} = \frac{\omega_1}{c} + \frac{\omega_1}{2c}\chi_{s1}^{(1)}$$
$$k_{s2} = \frac{\omega_2}{c}\sqrt{1+\chi_{s2}^{(1)}} = \frac{\omega_2}{c} + \frac{\omega_2}{2c}\chi_{s2}^{(1)} \tag{19}$$

The term $\chi_{s1}^{(1)}$ is the first-order linear susceptibility of Signal 1, its real part and imaginary part represent the dispersion and absorption respectively. This linear susceptibility is close to zero because there are few atoms populated at the level |2>, so the term $k_{s1} = \frac{\omega_1}{c}\sqrt{1+\chi_{s1}^{(1)}} \sim \frac{\omega_1}{c}$. The result of Fourier transform is close to the function $\delta(\tau-\tau')$.

$$\tilde{\Phi}(\tau - \tau') = \frac{1}{2\pi}\int d\omega_{s2} \Phi(\omega_{s2}) e^{-i\omega_{s2}(\tau-\tau')} \approx \delta(\tau - \tau'). \tag{20}$$

Then we obtain the simplified two-photon function shown below:

$$G^{(2)}(\tau) = \left| \frac{-iL\varepsilon_{p1}\varepsilon_{p2}\sqrt{\overline{\omega}_{s1}\overline{\omega}_{s2}}}{4\pi c} \int d\omega_{s2} \chi_{s2}^{(3)}(\omega_{s2}) e^{-i\omega_{s2}\tau} \right|^2. \tag{21}$$

We assume the detuning $\Delta_2 \sim 0$. In our experiment, $\Delta_1 \gg \gamma_1, \Gamma_1$. If we consider the Doppler effect in calculations as $\Delta_{1D} = \Delta_1 + \frac{\omega_{p2}}{c}v$, c is speed of light, v is the speed of atomic motion, then the two-photon function can be calculated with the following expression:

$$G^{(2)}(\tau) = \left| \int \frac{NLu_{21}u_{13}u_{34}u_{42}\varepsilon_{p1}\varepsilon_{p2}\sqrt{\overline{\omega}_{s1}\overline{\omega}_{s2}}}{4\hbar^3\varepsilon_0\pi c[i\Delta_{1D}(2\Gamma_1-\gamma_1)+\Omega_{p1}^2]} e^{-i\overline{\omega}_{s2}\tau}(e^{-\gamma_1\tau} - e^{-2\Gamma_1\tau - \frac{\gamma_1\Omega_{p1}^2}{\Delta_{1D}^2}\tau + i\frac{\Omega_{p1}^2}{\Delta_{1D}}\tau})\frac{1}{u\sqrt{2\pi}}e^{-v^2/u^2}dv \right|^2 \Xi(\tau) \tag{22}$$

where, $u = \sqrt{2k_B T/m}$, corresponding to the most probable velocity, $k_B$ is Boltzmann constant, $T$ is temperature, and $m$ means the mass of the Rb atom. The above equation describes the two-color correlation function in the time domain, which consists of two parts: oscillation term and decay term. When $\tau \geqq 0$, $\Xi(\tau) = 1$, and $\Xi(\tau) = 0$ for $\tau<0$. The decay term shows the width of correlation function, and the oscillation term describes oscillation phenomenon.

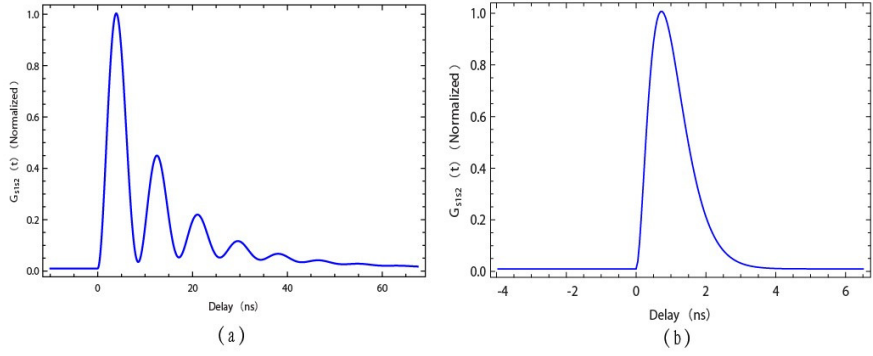

Fig. 3. (a) is the theoretical curve using the parameters $\gamma_1 = 3 \times 2\pi \times 10^6$ (s$^{-1}$), $\Gamma_1 = 1 \times 2\pi \times 10^6$ (s$^{-1}$), $\Omega_{p1} = 60 \times 2\pi \times 10^6$ (s$^{-1}$). Figure 3(b) is the theoretical curve describing the correlation function with large decay rates $\gamma_1 = 210 \times 2\pi \times 10^6$ (s$^{-1}$), $\Gamma_1 = 100 \times 2\pi \times 10^6$ (s$^{-1}$).

Using Eq. (22), taking the coefficients $\gamma_1, \Gamma_1$ as $3 \times 2\pi \times 10^6$(s$^{-1}$), $1 \times 2\pi \times 10^6$(s$^{-1}$) and $\Omega_{p1} = 60 \times 2\pi \times 10^6$(s$^{-1}$), $\Delta_1 = 25 \times 2\pi \times 10^6$(s$^{-1}$) (such parameters correspond to the resonance conditions in a cold atomic ensemble), we plot the correlation function shown in Fig. 3(a). The oscillation phenomenon is from the interference of two types of photon pairs generated through two different SFWM processes in atomic-gas media, can be explained by the dressed-state picture [25]. The oscillation of Fig. 3(a) is similar to the result shown in Ref [20], where a cold atomic ensemble is used. On the contrary, the oscillation disappears when large decay rates of $\gamma_1 = 100 \times 2\pi \times 10^6$ (s$^{-1}$), $\Gamma_1 = 210 \times 2\pi \times 10^6$ (s$^{-1}$), $\Delta_1 = 2.5 \times 2\pi \times 10^9$ (s$^{-1}$) are taken. Such parameters correspond to our experimental conditions. The numerical simulated result is shown in Fig. 3(b). We can see that the atomic decay rates affect the oscillation behavior, large decay rate induces the short dephasing time (of ~ns decay time), which causes the oscillations suppressed.

In order to describe our experiment more precisely, we consider the noise in detection. In our experiment, the main noise comes from the spontaneously emitted fluorescence at 780 nm. The number of the fluorescent photons at 1529 nm in our experiment is very small so that we can ignore them. The absorption and dispersion of a two level atom can be described $\chi^{(1)} = A / (\Delta_1 - i\gamma_1)$ [26]. The noise factor B can be written as:

$$B \sim \left| \int \frac{A}{(\Delta_{1D} - i\gamma_1)} \frac{1}{u\sqrt{2\pi}} e^{-v^2/u^2} dv \right|^2 \qquad (23)$$

In which, we define a coefficient of A, which describes the transition between |1> and |2>, is proportional to the rabi frequency of pump 2. The integral part is the Doppler averaging. The noise factor B becomes large with the increase of the intensity of Pump 2. The total correlation function containing noise can be described as

$$g_{s1s2}^{(2)}(\tau) = \frac{G^{(2)}(\tau) + B}{B} \qquad (24)$$

Figure. 4(a) and 4(b) are the theoretical results describe the relations between the cross-correlation and the intensity and detuning of Pump 2 respectively. From Fig. 4(a), we see that the small detuning would bring large noise in correlation measurement. When the detuning of Pump 2 is large, the noise is small, and $g_{s1s2}^{(2)}(\tau)$ becomes large too. This is very reasonable because with the increase of the detuning, the transition probability between the states |1> and |3> becomes small, therefore the possibility exciting two atoms or more between the states |1> and |3> at the same time becomes small too, it reduces the possibility of multi-photon emission, which is a main noise factor. Figure 4(b) is the calculated correlation vs the transition coefficient A. From that we could see that with the decrease of the power of the Pump 2, the $g_{s1s2}^{(2)}(\tau)$ becomes large. This is also reasonable because with the decrease of the power of the Pump 2, the possibility exciting two atoms or more between the states |1> and |3> at the same time becomes small too, it also reduces the possible multi-photon emission and the noise caused by it.

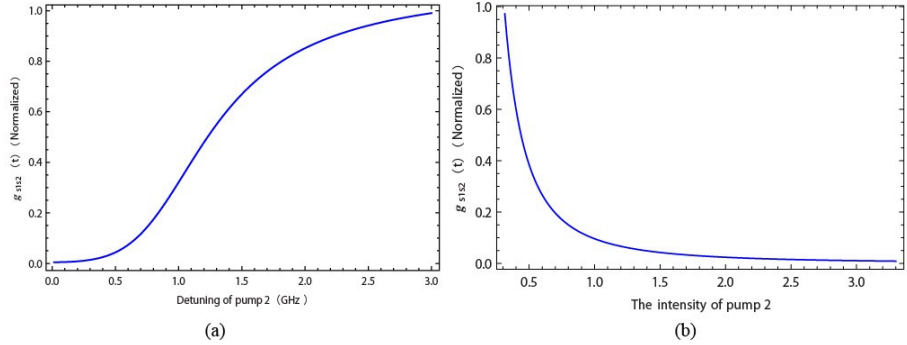

Fig. 4. (a) is the theoretical curve describing the correlation of biphoton vs the detuning of Pump 2. Figure 4(b) is the theoretical curve describing the relation between the correlation function and the power of Pump 2.

## 4. Experimental results

The 780 nm and 1529.4 nm lasers have 1 MHz bandwidths. By two-photon resonant absorption, we can fix the frequency of 1529.4 nm [27] with an accuracy 1 MHz determined by the laser bandwidths. The powers of Pump 1 and Pump 2 in our experiment are 16 mW and 55 $u$W respectively. The Rb cell is heated to $110°$, corresponding to the atomic density of $1.4 \times 10^{13}$ cm$^{-3}$. In order to obtain the cross-correlation function of the two correlated photons, we firstly perform a nonlinear four-waving mixing experiment for optimizing the alignment of experimental setup. After that, we perform the experiment for generating a photon pair by SFWM. The two pump beams Pump 1 and Pump 2 are input lights for producing non-classical correlated photon pairs. Correlated photons Signal 1 and Signal 2 in our experiment are almost counter-propagating. We collect the Signal 1 photons into a single-mode fiber with the efficiency of 50% and Signal 2 photons into a multi-mode fiber with the efficiency of 90% respectively. Through adjusting the delay of a delay generator DG535, we obtain the measurement of cross-correlation function with high accuracy shown in Fig. 5(a). The cross-correlation function is the convolution of our detector time bin width and DG535 time bin width. We scan DG535 with a step size of 0.1ns to measure the cross-correlation function (0.1ns step size is much smaller than the coherence time of two-photons and the resolution of detector ~ns). Therefore, the cross-correlation function is proportional to $G^{(2)}(\tau)t_c$, where $t_c$ is the step size of the delay generator. The dots are experimental data and the line is the theoretical fitting using the Eq. (24) with the parameters of $\gamma_1 = 210 \times 2\pi \times 10^6$ (s$^{-1}$), $\Gamma_1 = 100 \times 2\pi \times 10^6$ (s$^{-1}$). The full width at half maximum is about 1~2 ns, about the resolution of the detector we use. The short pulse width of the generated photon is due to the large decay of $\gamma_1$. The asymmetry of the cross-correlation is caused by the decay of the atoms. In our experiment, Signal 2 photon is always produced after Signal 1 photon. Therefore, the decay only goes to the positive direction of $t_{s2}-t_{s1}$. Please note that the line is scaled in order to fit the data. We could see that the experimental results are in reasonable agreement with the theoretical prediction.

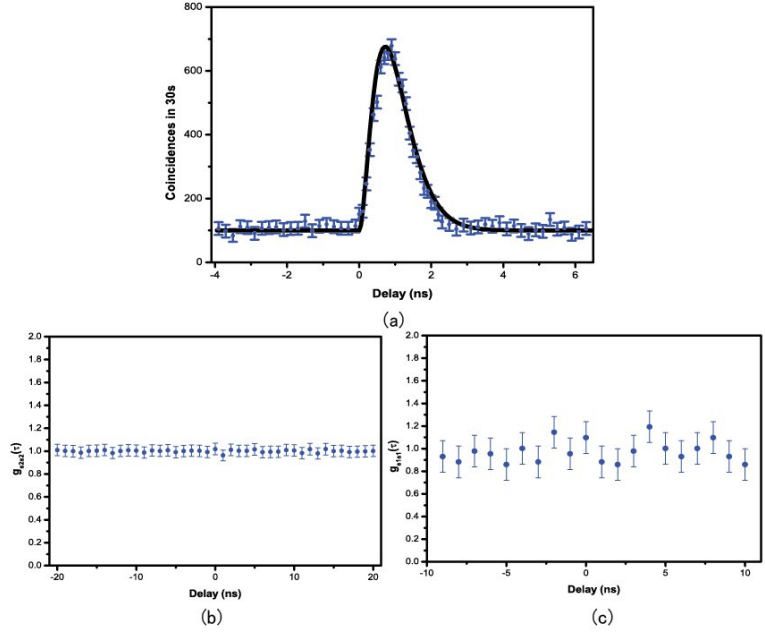

Fig. 5. (a) is the measurement of cross-correlation function of two signals. Black line is the theoretical fitted curve using Eq. (24). Figure 5(b) and Fig. 5(c) are the measurements of autocorrelation functions of Signal 2 and Signal 1 respectively.

Before measuring the autocorrelation function of Signal 1 or Signal 2 photons, we must align these autocorrelation coincidence circuits. In order to measure the autocorrelation function of Signal 1 photons, firstly, we use a strong light at 780 nm to pump a periodically poled potassium titanyl phosphate (PPKTP) crystal to generate a SPDC photon pair at 1560 nm. Then we measure the cross-correlation function between the correlated photon pairs (1560 nm) using two In-GaAs Photon Detectors (One In-GaAs photon detector works at internal trigger mode with 10 MHz frequency. The other detector works triggered from a delay pulse generated by the first detector when it fires.), to align the coincidence circuits for measuring autocorrelation of Signal 1 photons (1529.4 nm). Besides, we align the coincidence circuits for measuring autocorrelation function of Signal 2 by measuring autocorrelation of resonant fluorescence at 780 nm generated in atoms using two Si-APD detectors. After that, we begin to measure the autocorrelation functions of Signal 1 and Signal 2 photons. Figures 5(b) and 5(c) are the measured autocorrelation functions of Signal 2 photons and Signal 1 photons respectively. The single counts of Signal 1 and Signal 2 photons are 8000/s and 50000/s. The photon pair production rate M can be calculated by the formula $M = S_1 \times S_2/R_c$, which is $\sim 10^7$/s, where, $S_1$, $S_2$ are net single counts of Signal 1 and Signal 2, $R_c$ is the cross-coincidence count between Signal 1 and Signal 2. The estimated heralded single photon rate is about 280/s. The small $g_{s1,s2}(\tau)$ in our experiment causes the relatively small heralded single photon rate.

The non-classical correlation between the generated photons can be proved by checking whether the Cauchy-Schwarz inequality is violated. Usually classical lights satisfy the following equation [12]:

$$R = \frac{g_{s1,s2}(\tau)^2}{g_{s1,s1} g_{s2,s2}} \leq 1 \qquad (25)$$

where, $g_{s1,s2}(\tau)$, $g_{s1,s1}(\tau)$ and $g_{s2,s2}(\tau)$ are the cross-correlation and auto-correlations of the photons respectively. From the experimental data shown in Fig. 5(a), we obtain the cross-correlation $g_{s1,s2}(\tau)^2 = 46.3 \pm 2.7$. At the experimental temperature of Rb cell of 110°, the autocorrelation functions shown in Figs. 5(b) and 5(c) appear to be flat within the experimental uncertainty. This is due to the multiple scattering events in an optically thick medium at high operating temperature, which cause frequency redistribution of the photons across the entire Doppler width, leading to a very narrow autocorrelation. We could not resolve such narrow function with our detectors. When the temperature of the cell is decreased, the temporal extent of the correlation will become large, a peak can be observed. Such phenomenon has been analyzed theoretically in Ref. 21 and an autocorrelation peak at 780 nm has been experimentally obtained at low temperature in Ref. 22. The auto-correlations of Signal 1 and Signal 2 photons are $g_{s1,s1} = 1.00 \pm 0.04$ and $g_{s2,s2} = 1.00 \pm 0.14$, then we calculate R and

$$R = \frac{g_{s1,s2}(\tau)^2}{g_{s1,s1}g_{s2,s2}} = 48 \pm 12 > 1 \tag{26}$$

which is much large than 1, therefore the Cauchy-Schwarz inequality is strongly violated. The result of above inequality is a proof that the photons in a pair generated in our experiment are non-classical correlated. In our experiment, because of large noise existed in the degenerated ladder-type configuration, R is not very big. If we could reduce the noise further, a more large R could be obtained. In Ref. 22, a large $R = 495 \pm 56$ is obtained where a diamond configuration is used. In their experimental setup, four waves with large wavelength difference are used, therefore it is relatively easy to reduce the noise in their experiment. so they could obtain a quite large R. Another reason is from the use of a multimode fiber for collecting the Signal 2 photons. Single-mode collection for signal 2 is expected to give a better $g_{s1,s2}(\tau)$ since the two modes are expected to show higher correlations.

Keeping other parameters unchanged, we change the detuning of Pump 2 and measure the cross-correlation function of two signal photons shown in Fig. 6. The correlation $g_{s1,s2}(0)$ becomes larger with the increase of detuning of Pump 2, which is in agreement well with the theoretical prediction shown in Fig. 4(a).

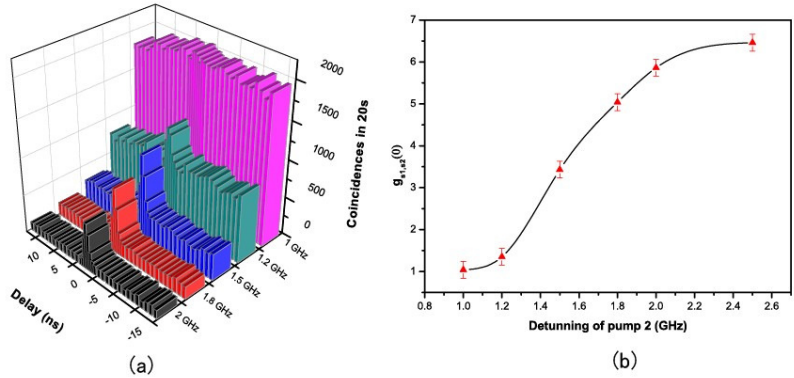

Fig. 6. (a) is the cross-correlation function vs the different detuningof Pump 2. Figure 6(b) is plot of the correlation $g_{s1,s2}(0)$ vs the detuning of Pump 2 from the data of Fig. 5(a) and Fig. 6(a). The solid line is the guide for eye.

Keeping the power of Pump 1 16.6 mW and the detuning of Pump 1 unchanged, we obtain results shown in Fig. 7. With the increment of the power of Pump 2, the correlation $g_{s1,s2}(0)$ becomes weak, shown in Fig. 7(a). This is also in agreement well with our theoretical prediction shown in Fig. 4(b). If the power of pump 2 is fixed to be 16$u$W, when we change the power of Pump 1, we find the correlation $g_{s1,s2}(0)$ becomes larger with the increase of the power of Pump 1, shown in Fig. 7(b). The reason is clear because with the increment of the power of the Pump 1, the transition probability between the states |3> to |4> becomes large. The atom on the state

|3> excited from the state |1> by Pump 2 can be excited to the state |4> with higher probability, which increase the probability of emitting an infrared photon and a visible photon directly.

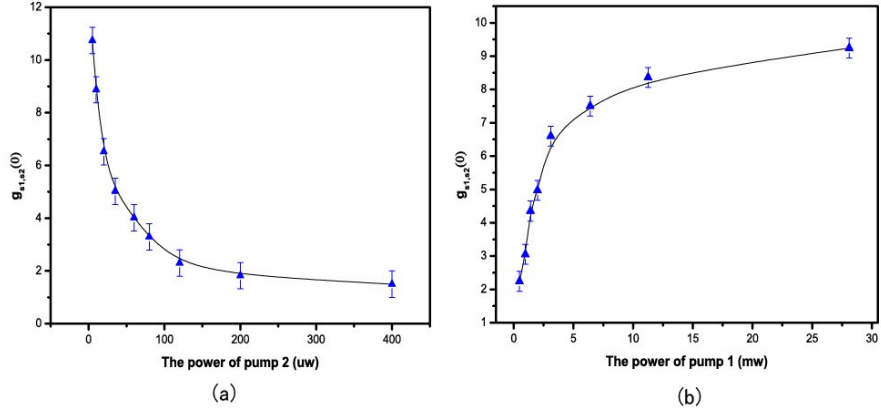

Fig. 7. (a) is the correlation $g_{s1,s2}(0)$ vs the power of Pump 2, and Fig. 7(b) is the correlation $g_{s1,s2}(0)$ vs the power of Pump 1. The solid line is the guide for eye.

## 5. Conclusion

In conclusion, we generate correlated two-color photon pairs using a ladder-type configuration in a hot $^{85}$Rb atomic vapor, and measure the cross-correlation function between the two-color photons. We obtain the strong violation of the Cauchy-Schwarz inequality. Besides, we give the relations between the power, single-photon detuning of pumps and the correlation functions. The theoretical calculations we give are in reasonable agreement with our experimental results. Our results are important to the research about quantum repeater based on atomic system.

### Acknowledgments

We thank Dr. Wei Chen and Dr. Bi-Heng Liu for kindly lending us single photon detector and for other technique support. This work was supported by the National Natural Science Foundation of China (Grants No. 10874171, No. 11174271), the National Fundamental Research Program of China (Grant No. 2011CB00200), the Innovation fund from CAS, Program for NCET.